\def\mode{0} 

\if 1\mode
    \documentclass[preprint,superscriptaddress,floatfix, nofootinbib]{revtex4-2}
    \usepackage{lineno}
    \linenumbers
    \usepackage{natbib}
\else
    \documentclass[twocolumn,superscriptaddress, aps, prx,floatfix, nofootinbib]{revtex4-2}
\fi

\usepackage{xcolor}
\usepackage{framed}
\definecolor{shadecolor}{RGB}{224,224,224}
\definecolor{orange(ryb)}{HTML}{FFA500}
\definecolor{lightorange(ryb)}{HTML}{FFB300}
\definecolor{dodgerblue}{HTML}{1E90FF}
\definecolor{lightdodgerblue}{HTML}{4dbff7}
\definecolor{crimson}{HTML}{FF4C4C}
\definecolor{pinkerton}{HTML}{e03185}
\definecolor{forest}{HTML}{6DD189}
\definecolor{lightishgray}{HTML}{DFDFDF}
\definecolor{error-red}{HTML}{EFB2B6}

\usepackage{dsfont}
\usepackage{graphicx}
\usepackage{dcolumn}
\usepackage{bm}
\usepackage{float}
\usepackage{siunitx}
\usepackage{braket}
\usepackage{amsmath,amssymb}
\usepackage[
	pdftex,
	colorlinks=true,
  	urlcolor=dodgerblue,
  	linkcolor=dodgerblue,
  	citecolor=dodgerblue
]{hyperref}

\usepackage{ulem} 
\begin{document}

\title{Topological Phase Transitions and Mixed State Order\\in a Hubbard Quantum Simulator}
\author{
Lin~Su$^{*,\dagger,1}$, 
Rahul~Sahay$^{*,1}$, 
Michal~Szurek$^{1}$, 
Alexander~Douglas$^{1}$, 
Ognjen~Markovi\'{c}$^{1}$, 
Ceren~B.~Dag$^{\S, 1,2,3}$, 
Ruben~Verresen$^{4}$, 
and Markus~Greiner$^{\ddagger,}$
}

\affiliation{
Department of Physics, Harvard University, Cambridge, Massachusetts 02138, USA\\
$^2$Department of Physics, Indiana University, Bloomington, Indiana 47405, USA\\
$^3$ITAMP, Harvard-Smithsonian Center for Astrophysics, Cambridge, MA 02138, USA\\
$^4$Pritzker School of Molecular Engineering, University of Chicago, Chicago, IL 60637, USA\\
$^*$These authors contributed equally, $^\dagger$ls4211@columbia.edu,
$^\S$cbdag@iu.edu,
$^\ddagger$greiner@physics.harvard.edu
}

\date{\today}
\begin{abstract}
Topological phase transitions challenge conventional paradigms in many-body physics by separating phases that are locally indistinguishable yet globally distinct.
Using a quantum simulator of interacting erbium atoms in an optical lattice, we observe such a transition between one-dimensional crystalline symmetry-protected topological phases (CSPTs).
We detect the critical point through non-local string order parameters and reveal its connection to the transition predicted between the Mott and Haldane insulators.
Moreover, we demonstrate a striking property: stacking two identical systems eliminates the transition, confirming the predicted group structure and invertibility of SPTs.
Finally, while introducing symmetry-breaking disorder also removes the transition, disorder averaging restores it.
Consequently, the adjacent phases realize a form of mixed-state quantum order wherein the criticality between them depends on the observer’s information.
Our results demonstrate how topology and information influence quantum phase transitions, opening the doors to probing novel critical phenomena in programmable quantum matter.
\end{abstract}

\maketitle

\section{Introduction} 

\begin{figure*}
    \centering
    \includegraphics[width=\textwidth]{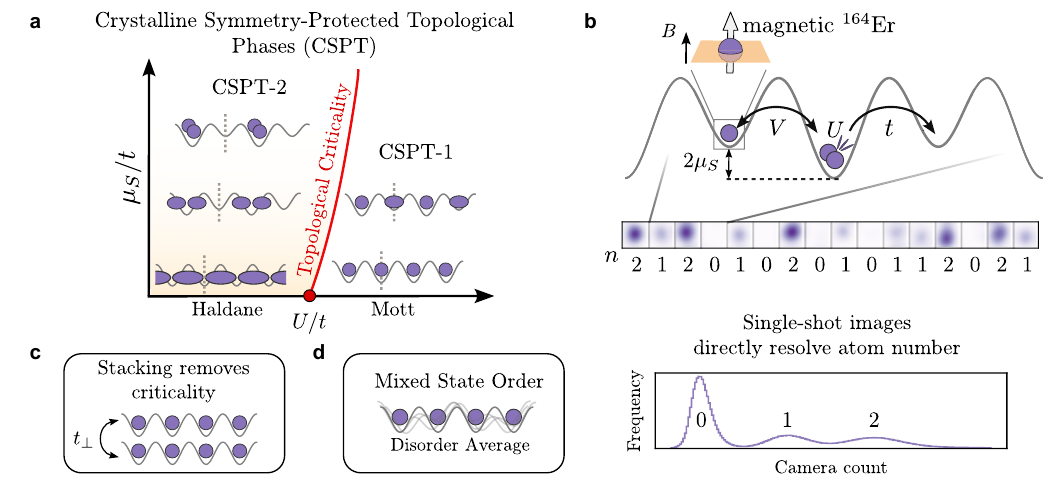}
    \caption{\textbf{Topological phases and criticality in a staggered dipolar Bose-Hubbard quantum simulator.}
    \textbf{a}, With long-range density-density interactions ($V$), a symmetry protected topological (SPT) phase, named the Haldane insulator (HI), emerges in the 1D extended Bose-Hubbard phase diagram \cite{Torre2006, Berg2008}.
    The HI is in a distinct phase from the Mott insulator (MI). Remarkably, the topological transition persists when a staggered chemical potential $\mu_S$ is introduced. In this regime, the HI and MI admit atomic limits which are in two distinct crystalline SPTs (CSPTs) protected by the parity conservation of bosons (Eq.~\eqref{eq: parity-generator}) and site-centered inversion symmetry (dashed line) \cite{Fuji2015}.
    \textbf{b}, Experimentally, we realize this model using bosonic $^{164}$Er atoms in an optical lattice. Particles tunnel with amplitude $t$ and experience on-site interactions of strength $U$. Dipolar interactions between erbium atoms induce long-range density-density coupling with nearest-neighbor strength $V$. A superlattice generates a staggered chemical potential $\mu_S$. We directly resolve the atom number per site without parity projection. A representative single-shot image is shown, with the digitized filling below. The camera-count histogram displays peaks at $0$, $1$, and $2$ particles per site.
    \textbf{c}, We probe the stability of the topological criticality by stacking two instances of the 1D chain, demonstrating that these quantum phases of matter are invertible.
    \textbf{d}, Moreover, we demonstrate that the criticality is unstable to programmable symmetry-breaking disorder, but is restored when averaging over disorder. In this sense, these SPTs correspond to distinct mixed-state quantum phases of matter~\cite{Ma2023}.
    }
    \label{fig: setup}
\end{figure*}

Systems of many interacting particles in equilibrium can undergo phase transitions where macroscopic properties abruptly change~\cite{Sachdev1999}. 
Traditionally, these transitions are described by the Landau paradigm, which posits that transitions occur when the correlations of local observables (or ``order parameters'') become long-range and spontaneously break the system's symmetries.
This theory successfully explains many finite-temperature transitions, such as between liquids and crystals, or between different types of magnets.
However, at near-zero temperatures, quantum effects lead to ``topological'' phase transitions with no temperature-driven analog, where nonlocal quantum correlations separate phases that are indistinguishable by local measurements.

Remarkably, such topological transitions have been predicted even for one-dimensional quantum systems, namely between symmetry-protected topological (SPT) phases~\cite{Senthil2015, Pollmann2010, Turner11, Fidkowski11, Pollmann2012, Schuch2011, Chen2011, Chen2012}.
Recent advances in analog quantum simulation have enabled the realization of such 1D SPT phases---in platforms ranging from ultracold atoms~\cite{Atala2013, Meier2016, Sylvain2019, Sompet2022}, superconducting circuits~\cite{Cai2019}, and trapped ions~\cite{Katz2024}---in addition to digital simulations \cite{Choo2018, Smith2022, Herrmann2022}.
However, detecting the quantum critical points between these phases has remained an outstanding challenge in analog simulators, requiring a tunable parameter to sweep across the transition.

In this work, we realize such quantum critical points between SPT phases in an analog quantum simulator (Fig.~\ref{fig: setup}a) and demonstrate their topological nature by probing their stability and instability under stacking and disorder (Fig.~\ref{fig: setup}c, d). We employ a quantum simulator \cite{Bloch2008, Gross2017, Bohrdt2021} to realize a variant of the Bose–Hubbard model featuring long-range interactions and a staggered chemical potential (Fig.~\ref{fig: setup}b):
\begin{equation}
\label{eq: Hamiltonian}
\begin{split}
H=-t\sum_{x}(\hat{a}_x^\dag \hat{a}_{x + 1}+\mathrm{h.c.})+\frac{U}{2}\sum_x\hat{n}_x(\hat{n}_x-1) \\
+\sum_{x, r>0}V_{r}\hat{n}_x\hat{n}_{x + r} - \mu_S\sum_x(-1)^x\hat{n}_x.
\end{split}
\end{equation} 
Here, $t$, $U$, and $\mu_S$ denote the tunneling, on-site interaction, and staggered potential strength, respectively; $a_x^{\dagger}$ ($a_x$) creates (annihilates) a boson at site $x$, and $\hat{n}_x = a_x^{\dagger} a_x$ is the number operator. $V_r$ describes long-range dipolar interactions at distance $r$ (see Supplementary Materials, SM). We refer to this model as the staggered dipolar Bose-Hubbard model.
Previous work \cite{Torre2006,Berg2008} has predicted a Haldane insulator \cite{Haldane1983} in its phase diagram, which we also probe.

To simulate this Hamiltonian, we use magnetic erbium atoms with dipolar interactions~\cite{Su2023} and tune the on-site interaction via Fano-Feshbach resonances~\cite{Chin2010} (Fig.~\ref{fig: setup}b). We adiabatically load bosonic $^{164}$Er atoms into 2D optical lattices with deep confinement in one direction to create isolated 1D chains (SM). Thousands of experimental realizations are performed, followed by data selection procedures (SM) that include restricting the analysis to samples with unit filling. The boundary conditions are described in the SM. Site-resolved imaging~\cite{Bakr2010, Sherson2010} then yields atom-number measurements without parity projection~\cite{Su2024}.

\section{Crystalline SPTs and Their Transitions} \label{sec:CSPT-transitions}

\begin{figure*}
    \centering
    \includegraphics[width=\textwidth]{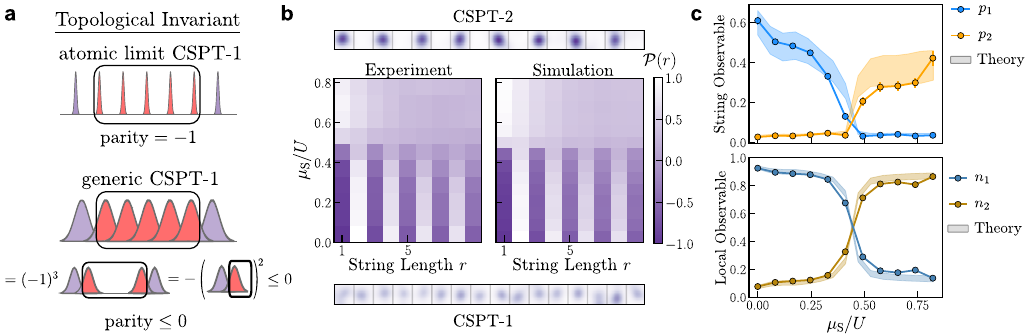}
    \caption{\textbf{Quantum phase transition between crystalline symmetry protected topological phases.} We investigate the quantum phase transition between two CSPTs in one-dimensional chains of 16 sites. \textbf{a}, For the atomic limit CSPT, the parity of an odd-length string for a unity-filled state has $-1$ parity.
    Even if quantum fluctuations, e.g., induced by tunneling, take us away from this atomic-limit state and the bosons mildly delocalize, the parity remains negative in the presence of site-centered inversion symmetry. \textbf{b}, The sign of the parity string of odd length distinguishes between the two CSPTs and provides a topological invariant.
    As $\mu_S$ crosses the quantum critical point, the system transitions from the CSPT-1 phase (Mott insulator-like, bottom) to the CSPT-2 phase (Charge Density Wave-like, top).
    Our experimentally measured parity string shows good agreement with the results from simulations taking into account readout infidelities (SM).
    \textbf{c}, We use the parity string measurements to define two effective ``string order parameters'' for CSPT-1 and 2 ($p_1$ and $p_2$) shown in the top panel.
    The experimental observables are plotted with error bars (bootstrapping standard error) and compared to simulations (shades indicate 95\% confidence interval in this entire paper).
    On the CSPT-1 side of the transition, the $p_2$ observable (orange) is effectively zero (limited by finite imaging fidelity), while the local average density observable $n_2$ (brown) is significantly non-zero, highlighting the topological nature of the CSPTs. This shows that local observables cannot differentiate between the two CSPTs, whereas the parity ``string order parameter'' can. The crossover of the local observables is not due to finite-size effects (SM).
    }
    \label{fig: CSPT_ground}
\end{figure*}

We begin by highlighting two symmetries protecting the SPT phases and their transitions.
First, due to particle number conservation, the system preserves the total \textit{parity} (even or oddness) of the number of bosons in the chain.
This yields a $\mathbb{Z}_2$ symmetry of the Hamiltonian generated by: 
\begin{equation} \label{eq: parity-generator}
 \hat{\mathcal{P}} = \prod_{x} e^{i \pi \hat{n}_x}   .
\end{equation} 
In addition, the model exhibits a site-centered inversion symmetry $\hat{\mathcal I}$: a \textit{crystalline} symmetry that inverts the lattice about a given site.
The topological phases we study are protected by this crystalline symmetry as well as the internal parity symmetry, and can consequently be referred to as crystalline SPT phases (CSPTs) \cite{Fu11, Turner10inversion, Hughes11, Fuji2015, Song17}.

These phases arise in the phase diagram of our system at unit filling $\bar{n} =1$ and with a large on-site Hubbard interaction compared to tunneling, as shown in the upper section of the phase diagram in Fig.~\ref{fig: setup}a.
The first CSPT phase (CSPT-$1$) is essentially the Mott insulator, which appears for small values of the staggered chemical potential.
Here, the strong on-site Hubbard interaction stabilizes an average number of bosons on every site close to $1$.
In contrast, the second CSPT (CSPT-$2$) is found when the staggered chemical potential is large, and consequently, there is an average of two bosons on the odd sites (`doublons') and no bosons on the even sites.
These configurations are schematically depicted in the phase diagram of Fig.~\ref{fig: setup}a, where the classical states ($\ket{1111\cdots}$ and $\ket{2020\cdots}$) become exact in the atomic limit, i.e. when the tunneling $t\to 0$.

These two states cannot be adiabatically connected in the presence of the aforementioned parity $\hat{\mathcal P}$ and inversion $\hat{\mathcal I}$~\cite{Fuji2015}. 
Indeed, while there exists no local order parameter to distinguish these two phases in the presence of quantum fluctuations \cite{Fuji2015}, they are instead differentiated by a nonlocal topological invariant.
This invariant is precisely the \textit{sign} of the average boson parity in a large region of odd length.
In particular, this sign is negative (positive) in CSPT-1 (CSPT-2), and cannot change without closing the energy gap (Fig.~\ref{fig: CSPT_ground}a and SM).

\begin{figure*}
    \centering
    \includegraphics[width=\textwidth]{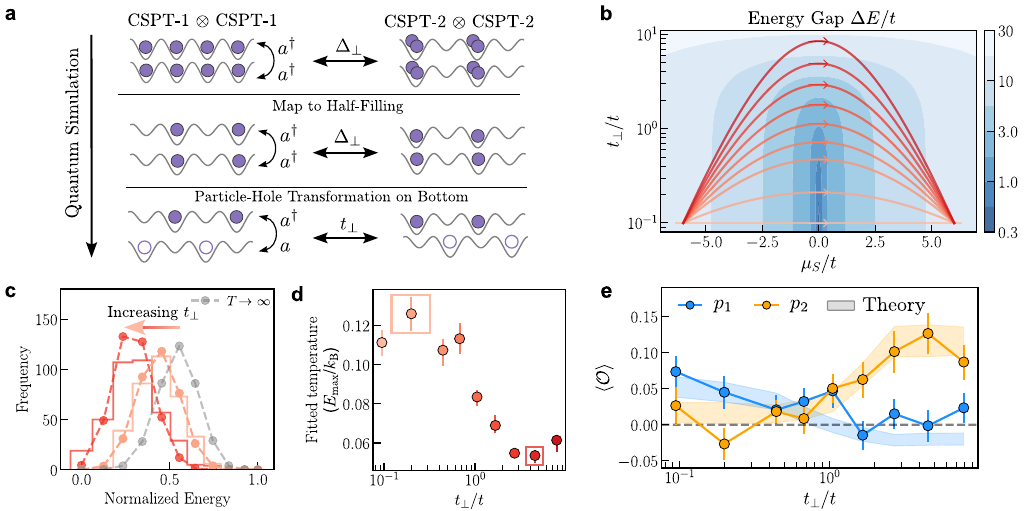}
    \caption{\textbf{Elimination of the phase transition via stacking.} Coupling an even number of $\mathbb{Z}_2$ topological chains can trivialize the topology. \textbf{a}, To quantum simulate the pair-creation coupling in Eq.~\eqref{eq: coupling}, we map our model (top row) to a half-filling representation (middle row), and then apply a particle-hole transformation to the bottom chain (bottom row). Our system size is 2 by 10 sites. \textbf{b}, When inter-chain tunneling $t_\perp$ is nonzero, the numerically computed energy gap at the critical point of a single CSPT chain (e.g.,~$\mu_S/t = 0$) becomes finite. To probe this gap, we initialize two CSPT-1 chains at minimal $t_\perp$ and ramp into the CSPT-2 regime via various intermediate values. We vary the staggered chemical potential $\mu_S$ linearly in time while simultaneously ramping $t_\perp$ to a final configuration with negligible coupling, and read out snapshots of the resulting state. \textbf{c}, At the final $\mu_S = 6t$ point, tunneling is negligible, and the density of states can be approximated classically. We analyze each snapshot to compute energy histograms of the final state. As $t_\perp$ increases, the extracted temperature decreases. Colored bars show data; colored dotted lines are Gibbs distribution fits; the gray dotted line shows the infinite-temperature distribution. \textbf{d}, Measured temperatures (error bars: central 68\% percentile via bootstrap) systematically decrease with increasing $t_\perp$. \textbf{e}, Parity order parameters versus the maximum $t_\perp$ (error bars: bootstrapped standard error) reveal a crossover into the CSPT-2 phase only at sufficiently large coupling, consistent with the presence of a gapped path.
    }
    \label{fig: Coupled_CSPT}
\end{figure*}

To demonstrate the two different topological phases in our quantum simulator, we experimentally measure \cite{Hilker17, Wei23} this topological invariant via the parity ``string operator'' defined as:
\begin{equation}
\label{eq: POP}
\mathcal{P}_{x_0}(r) =  \left\langle \prod_{x=x_0}^{x_0+r} e^{i\pi \hat n_x}\right\rangle \equiv \langle \hat{\mathcal{P}}_{x_0}(r) \rangle,
\end{equation}
which precisely measures the average parity of bosons in a region.
The experimental results are shown in Fig.~\ref{fig: CSPT_ground}b as a function of the staggered field over the interaction strength, $\mu_S/U$.
As expected, for weak staggered fields, the parity is negative for odd-length strings---reflecting the topological invariant---and positive for even-length strings.
For strong staggered fields, the parity is positive for all string lengths.
These observations are consistent with numerical simulations (SM).

The sharp distinction between these phases can be diagnosed by computing the staggered and uniform parity ``string order parameters" $p_1$ and $p_2$, defined by averaging $(-1)^{r} \mathcal{P}_{x_0}(r)$ and $\mathcal{P}_{x_0}(r)$ respectively over sufficiently long $r$ (SM).
These order parameters are shown in the top panel of Fig.~\ref{fig: CSPT_ground}c. They clearly distinguish the two phases, being either finite or vanishing, with a sharp transition at the critical point.
One way to argue that these order parameters vanish outside of their respective phases is by using an approximate particle-hole symmetry of the model in this triplon-free regime (SM).
In contrast, we also define local observables of the average density fluctuations defined as $n_2= \langle (-1)^i\delta\hat{n}_i\rangle $, where $\delta \hat{n}_i=\hat{n}_i-1$, and $n_1= 1 - n_2$ (Fig.~\ref{fig: CSPT_ground}c bottom). Their smooth variation across the phase transition reflects the inability of local observables to faithfully distinguish between the two phases --- a defining feature of topological phases (SM). These observations strongly support the existence of two distinct CSPTs at $\mu_S \ll U$ and $\mu_S \gg U$, separated by a quantum phase transition, in agreement with our large-scale ground state numerical simulations~\cite{Sahay2025} (SM).

\section{Coupling an even number of chains removes the topological transition}

After identifying the quantum phase transition between the CSPTs, we now show how this transition can be removed by leveraging a key property of SPTs.
That is, by stacking and coupling a certain number of identical chains, one can render the overall state topologically trivial \cite{Turner11, Fidkowski11, Pollmann2012, Schuch2011, Chen2011, Chen2012}. This allows for a continuous, gapped path between distinct phases---removing the intervening quantum critical point.
Such a property is in sharp contrast to symmetry-breaking phases, where the stack of multiple such states remains symmetry-breaking and consequently cannot be connected back to symmetry-respecting states.

We probe this stacking property by adding the following pairing term to the two-chain version of Eq.~\eqref{eq: Hamiltonian}
\begin{equation}
    H_{\perp} = \Delta_{\perp} \sum_{x} a^{\dagger}_{\mathsf{T}, x} a^{\dagger}_{\mathsf{B}, x}   + \text{h.c.},
\label{eq: coupling}
\end{equation}
which results in a gapped path between the stacks of two CSPT-$1$ and CSPT-$2$ phases (Fig.~\ref{fig: Coupled_CSPT}a, top). Here, $a_{x, \mathsf{T}/\mathsf{B}}^{(\dagger)}$ labels the bosonic annihilation (creation) operator in the top/bottom layer.
Crucially, the above term preserves the protecting $\mathbb{Z}_2$ parity symmetry generated by Eq.~\eqref{eq: parity-generator}, as well as site-centered inversion.

We simulate the pairing term using a particle-hole mapping, as illustrated in Fig.~\ref{fig: Coupled_CSPT}a and detailed in the SM. Starting deep in the CSPT-1 states of both chains with negligible $t_\perp$, we ramp across the critical point of a CSPT chain,~$\mu_S=0$, into a regime where the ground state of each decoupled chain is the CSPT-2 state before imaging. 
The numerical contour plot in Fig.~\ref{fig: Coupled_CSPT}b shows that, with increasing $t_\perp$ between two chains, the many-body energy gap between the ground and the first excited states increases at $\mu_S=0$, removing the quantum criticality. The ramps in $\mu_S$ and $t_\perp$ follow the differently shaded red lines at a constant rate for the $\mu_S$ ramp.

At low interchain tunneling $t_\perp \rightarrow 0$, the ramp cannot be adiabatic as we cross the vanishing energy gap at the quantum critical point. The system heats up, resulting in low fidelity of the final state with respect to the CSPT-2 ground state.
This is revealed by measurement of the CSPT-2 order parameter $p_2$ (Fig.~\ref{fig: Coupled_CSPT}e), which is much smaller than the ground state expectation for small $t_{\perp}$.
Similarly, the CSPT-1 order parameter does not vanish in this regime. In contrast, at larger values of $t_\perp$, even though the total path traveled in the phase diagram is longer---because we have to change $t_\perp$ while simultaneously changing $\mu_S$ for the same total sweep time---we observe a more faithful preparation of CSPT-2 state as seen by a larger value of its order parameter.
\begin{figure*}
    \centering
    \includegraphics[width=\textwidth]{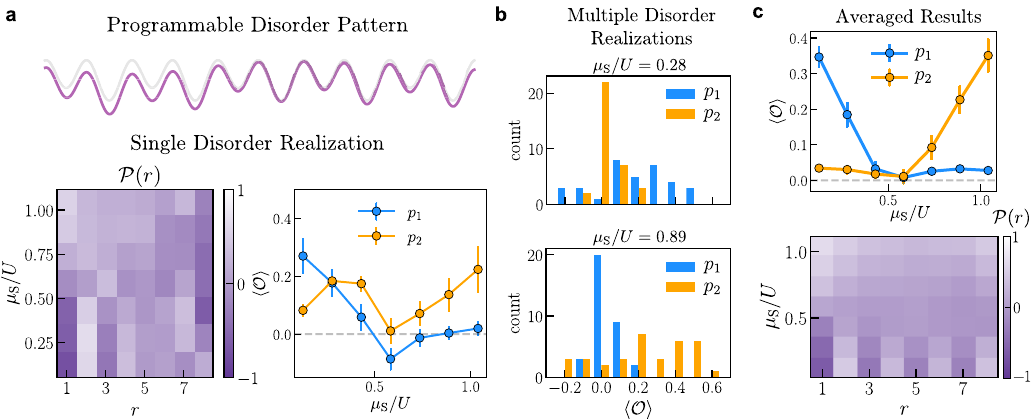}
    \caption{\textbf{Mixed-state CSPT from local disorder.} \textbf{a}, To demonstrate the importance of site-centered symmetry for protecting the CSPTs, we intentionally add different programmable chemical potential disorder, changing the original light gray lattice to the purple lattice. In the lower panel, the order parameters of a single chemical potential disorder realization (bootstrapping standard error) no longer show clear and distinct phases. The left side lower panel shows the order profile in space for different staggered potentials, while the right side lower panel integrates over the space to plot string order parameters. \textbf{b}, The histograms of the distribution of the $p_1$ and $p_2$ are plotted for the different disorder realizations. The disorder causes strong deformation in the order parameters for each disorder realization. \textbf{c}, Since the CSPTs are protected by average symmetry, once we average over the different disorder realizations, we obtain a clean signal of two distinct mixed-state phases (standard error of the mean) as seen from the order parameters in the top panel.
    The order parameters do not reach exactly zero due to finite imaging fidelity, finite size effects, and the degree of approximation of the particle-hole symmetry (SM), but still can distinguish between the two phases.
    The topological invariant is precisely the sign of the averaged parity string of odd length, seen to be positive for large $\mu_S$ and negative for small $\mu_S$ in the bottom panel.}
    \label{fig: CSPT_average}
\end{figure*}

The larger energy gap allows us to ramp across the phase diagram with less heating, which is also reflected in the final temperature of the prepared CSPT-2 states (Fig.~\ref{fig: Coupled_CSPT}c).
The energy of the final state can be well approximated by the energy of the classical state because, at the readout stage, the intrachain tunneling ($t$) is much smaller than other energy scales ($\mu_S$, $U$, and $V$). Based on each snapshot, we compute the energy of the final state and obtain a histogram of the state's distribution.
We then fit a temperature and observe the shift towards lower temperature, as $t_\perp$ is increased.

This provides experimental evidence that coupling two CSPT chains opens the energy gap and eliminates the quantum critical point, as predicted for $\mathbb{Z}_2$ topological phases~\cite{Gu2009}. This also demonstrates the `invertibility' of SPT phases~\cite{Freed2014}---a defining feature that sets them apart from non-invertible phases such as symmetry-breaking order or intrinsic topological order.

\section{Mixed State Order from Average Symmetry}

Thus far, we have demonstrated the presence of two distinct CSPTs in our experiment and exhibited a hallmark feature of the transition between them, i.e.,~the disappearance of the transition upon stacking.
Crucially, our results depended on the presence of both protecting symmetries of the CSPT---the crystalline site-centered inversion symmetry and the boson parity symmetry---as seen by the fact that the invariant was computed from a parity string of a site-centered inversion symmetric region (namely, one of odd length).
To further substantiate our claim that crystalline site-centered inversion symmetry is essential, we explicitly break this symmetry by implementing a programmable chemical potential disorder pattern shown at the top of Fig.~\ref{fig: CSPT_average}a, whose magnitude is comparable to the largest term in the Hamiltonian (peak-to-peak $\approx U$, see SM).

In our experiment, we adiabatically prepare the ground state of the disordered staggered dipolar Bose-Hubbard model $t \ll U, \mu_S$ by ramping down the tunneling from a superfluid while ramping up the chemical potential disorder pattern.
For any given realization of the disorder, we repeat the experiment thousands of times (SM) to sample the wavefunction.
This yields the parity string order parameters $p_1$ and $p_2$ shown in Fig.~\ref{fig: CSPT_average}a bottom, which show no signature of a transition with varying $\mu_S$, indicating that the transition is removed when inversion symmetry is broken.

However, let us now consider \textit{averaging} over disorder realizations (SM). 
We find that the transition is highly sensitive to whether or not we use the knowledge of the disorder.
Indeed, we found that when the disorder pattern is known, the transition disappears.
However, if this knowledge is erased through averaging---and hence the system forms a mixed state ensemble $\rho$ of the different disordered ground states (SM)---the transition re-emerges \cite{Ma2023}.

Although for each realization there is no sharp distinction between regimes $\mu_S \ll U$ and $\mu_S \gg U$ (as seen by the spread in parity values present in the histograms in Fig.~\ref{fig: CSPT_average}b), upon averaging over disorder, a sharp distinction is restored (Fig.~\ref{fig: CSPT_average}c). 
This is because the crystalline SPTs we study are also  \textit{average SPTs}, i.e.,~topological phases of the mixed states that arise due to the randomness of the disorder \cite{Fu2012, Ringel12, Mong12, Fulga14, Ma2023, Ma23_b, De2022, Coser2019, Fan2024, Bao2023, Verstraete2009, Diehl2008, Sang2024, Sang2025}.
Such average SPTs can be shown in theory to be protected by the $\mathbb{Z}_2$ parity symmetry along with an \textit{average} site-centered inversion symmetry of the mixed state corresponding to $\hat{\mathcal{I}} \rho \hat{\mathcal{I}}^{\dagger} = \rho$, where $\hat{\mathcal{I}}$ is the unitary operator implementing the action of inversion.
Our observations suggest that symmetry protection in realistic quantum systems can persist even in the presence of symmetry-breaking disorder, opening up the exploration of mixed-state phases.

\section{The Haldane-to-Mott Insulator Transition \label{sec:HI}}

\begin{figure}
    \centering
    \includegraphics[width=247pt]{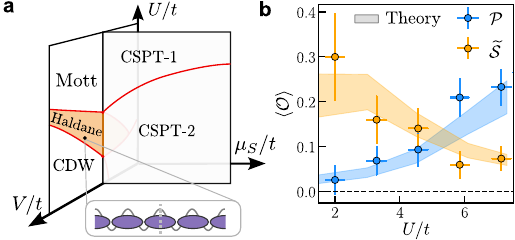}
    \caption{\textbf{Haldane-to-Mott insulator transition.} A well-known SPT phase in the dipolar Bose-Hubbard model is the Haldane Insulator (HI), which can be connected to the CSPT-2 phase. \textbf{a}, Once the staggered field is turned off, the CSPT-1 and -2 phases become Mott and Haldane insulators in the presence of off-site interactions ($V$). A schematic wavefunction for the HI described in Eq.~\eqref{eq:model-wavefunction} is shown in the inset. \textbf{b}, We tune the on-site interaction $U$ without the staggered field $\mu_S$ at fixed $V/t\approx3$. The order parameters show good agreement with simulated results. The vertical error bars are obtained from bootstrapping standard error; the horizontal error bars are estimated standard error.
    }
    \label{fig: HI}
\end{figure}

We conclude by presenting evidence for the Haldane SPT phase transition in our system and elucidating its connection to the CSPT transition explored in this work.
The dipolar Bose–Hubbard model is predicted to host a Mott insulator (MI), separated from the Haldane insulator (HI)~\cite{Haldane1983, Buyers1986, Xu1996, Torre2006} by a quantum phase transition when crystalline \textit{bond-centered} inversion and parity symmetries are present at $\mu_S = 0$, as shown in Fig.~\ref{fig: HI}a. The HI is a CSPT phase characterized by protected quantum entanglement \cite{Pollmann2012}.

Naturally, CSPT-1 and MI correspond to the same boson pattern and hence the same phase of matter. More intriguingly, the CSPT-2 phase we encountered for $\mu_S \neq 0$ adiabatically connects to HI as we take $\mu_S \to 0$ for suitable $U$ and $V$ \cite{Fuji2015}.
The HI can be viewed as a state where bosons are delocalized across lattice bonds, making it symmetric under parity as well as site- and bond-centered inversion symmetries (Fig.~\ref{fig: HI}a; see SM for the model wave function).
If we turn on the staggering $\mu_S \neq 0$, these delocalized bosons gradually move together to form the CSPT-2 state as schematically shown in Fig.~\ref{fig: setup}a.
Moreover, the direct topological phase transition between CSPT-1 and CSPT-2 (for $\mu_S \neq 0$) is in the same universality class as the transition between MI and HI (for $\mu_S = 0$) \cite{Sahay2025}.
We can thus explore the same topological phase transition in two very different energetic regimes.

The HI phase is known to exhibit the following string order parameter~\cite{Nijs1989, Pollmann2012Detection}
\begin{equation}
\label{eq: SOP}
\mathcal{S}_{x_0} (r)=\left\langle\delta \hat n_{x_0 -1} \hat{\mathcal{P}}_{x_0}(r)\delta \hat n_{x_0+r + 1}\right\rangle.
\end{equation}
This is similar to the parity string operator used for CSPT-1 and -2, but with the addition of endpoint dressing $\delta n = n - 1$. This modification ensures that Eq.~\eqref{eq: SOP} remains nonzero, and its positive value is consistent with CSPT-2 phenomenology (SM).
The hallmark signature of the Haldane insulator is then a contrast between the long-range order of $\mathcal{S}_{x_0}(r)$ along with the parity string operator $\mathcal{P}_{x_0}(r)$ decaying to zero at long distances.

To study the transition between the HI and MI, we turn off the staggered potential and adiabatically load our atoms into the lattice while setting magnetic fields to probe different on-site interactions $U$~\cite{Patscheider2022}.
Since snapshots where $\delta n = 0$ limit the magnitude of Haldane string order parameter $\mathcal{S}_{x_0}(r)$ and it oscillates as a function or $r$, we normalize it as $\widetilde{\mathcal{S}} =  (-1)^{r}\eta \mathcal{S}$ with $\eta=1/\langle |\delta n_i||\delta n_{i+r}|\rangle$~\cite{Sompet2022} computed for the state with the maximum string order parameter in the probed parameter space ($V/t \approx 3$).
The result is shown in Fig.~\ref{fig: HI}b and reveals two regimes of the phase diagram, one with Haldane-like correlations and the other with Mott-like correlations.
In our finite-size system (10 sites), the order parameters are not expected to exhibit sharp features due to large correlation length (SM).
Nevertheless, the order parameters we observe are consistent with the simulated results for our finite-size system.

\section{Conclusion and outlook}

Our study reveals that the topological criticality at the transition between the Haldane insulator (HI) and the Mott insulator (MI) can be accessed when probed between crystalline symmetry-protected topological (CSPT) phases.
This underscores the power of tunable quantum simulators, which grant access to a broad parameter space and enable the study of quantum criticality in otherwise inaccessible regimes. While local observables fail to distinguish these phases away from the atomic limit, we identified nonlocal string order parameters for the HI-MI and CSPT transitions. Both transitions are predicted to exhibit stacking behavior, which defines the group structure of SPTs and affirms their invertible nature. We demonstrated this property through dynamic ramping across the CSPT critical point. In the CSPT setting, we also observed the emergence of mixed state order, which generalizes topological invariants into mixed states.

Our uncovering of quantum phase transitions between 1D phases admitting atomic limits raises the possibility of analogous phenomena in higher dimensions---an open direction for both theoretical and experimental exploration. These transitions may help chart adiabatic pathways between topologically trivial and non-trivial phases, motivating novel approaches to the discovery of phases of matter~\cite{Semeghini2021} starting from experimentally accessible product states such as paramagnets~\cite{Scholl2021, Ebadi2021}. In parallel, our demonstration of mixed-state order raises compelling questions about its stability, the nature of intervening phases, and how mixed-state perspectives might refine our understanding of real-world quantum systems. Concretely, this same approach could be used to realize topological phases with average crystalline symmetries which can only exist in the presence of disorder \cite{Ma23_b,intrinsicACSPT}. Moreover, it would be interesting to compare the state preparation fidelity of SPT phases via explicit symmetry breaking versus the stacking property demonstrated in our work. The latter might be especially useful for preparing two-dimensional SPT phases protected by, e.g., particle parity symmetry \cite{LevinGu}, which would be difficult to break explicitly.
Our results emphasize that even seemingly simple models can harbor nontrivial topology, reinforcing the idea that topological tools are not just abstract concepts but practical frameworks for identifying and organizing quantum phases.

\vspace{4pt}

\textbf{Note added.} During the finalization of this manuscript, we became aware of a preprint on mixed state order in a digital quantum device~\cite{Zhang2025} and a preprint on average SPT in a Rydberg tweezer quantum simulator~\cite{Yue2025} which appeared in the same month as the present preprint.

\section*{Supplementary Materials}

\subsection*{Experimental details}

\subsubsection*{Model calibration}

On-site interactions ($U=U_\textrm{s}+U_\textrm{dd}$) are measured by modulating the lattice intensity. We load a unity-filling Mott insulator, modulate the lattice intensity, hold for a duration comparable to the doublon lifetime, and finally measure the atom numbers. The resonance frequency $\nu$ in the Mott-insulator regime is given by $h\nu\approx U_\textrm{s}+U_\textrm{dd}-V_\textrm{nn}$~\cite{Chomaz2016} ($V_{nn}$ is along the direction where the lattice is modulated). We calculate the on-site ($U_\textrm{dd}$) and off-site ($V$) dipolar interaction contributions, taking into account the finite Wannier(-Stark) functions. Our model agrees well with the experimental measurements for different dipolar angles and lattice depths. The off-site dipolar interaction ($V$) does not strictly follow $1/r^3$ decay due to the finite Wannier function size~\cite{Wall2013}. Instead, the first few nearest neighbors experience interactions that decay roughly as $1/r^{2.7}$, which does not modify the essence of the phases studied here. Tunneling energy ($t$) is calculated after measuring the lattice depth via intensity modulation. The staggered chemical potential ($\mu_S$) is calibrated via intensity modulation. The intrinsic (non-programmable) chemical potential disorder without the accordion lattice is estimated to be $\pm h\times$ 2.5~Hz, which is almost 4 times smaller than the smallest energy scale written explicitly in the Hamiltonian ($t$) and sufficiently small for the realization of the Haldane-Insulator-like states~\cite{Deng2013, Lv2018}. The super-exchange energy ($t^2/U$) when realizing the Haldane insulator is roughly 3~Hz and almost negligible when realizing CSPTs. The density-induced tunneling is -0.6~Hz~\cite{Baier2016}, which is more than an order of magnitude smaller than the tunneling, so we ignore it.

\begin{figure}
    \centering
    \includegraphics[width=0.48\textwidth]{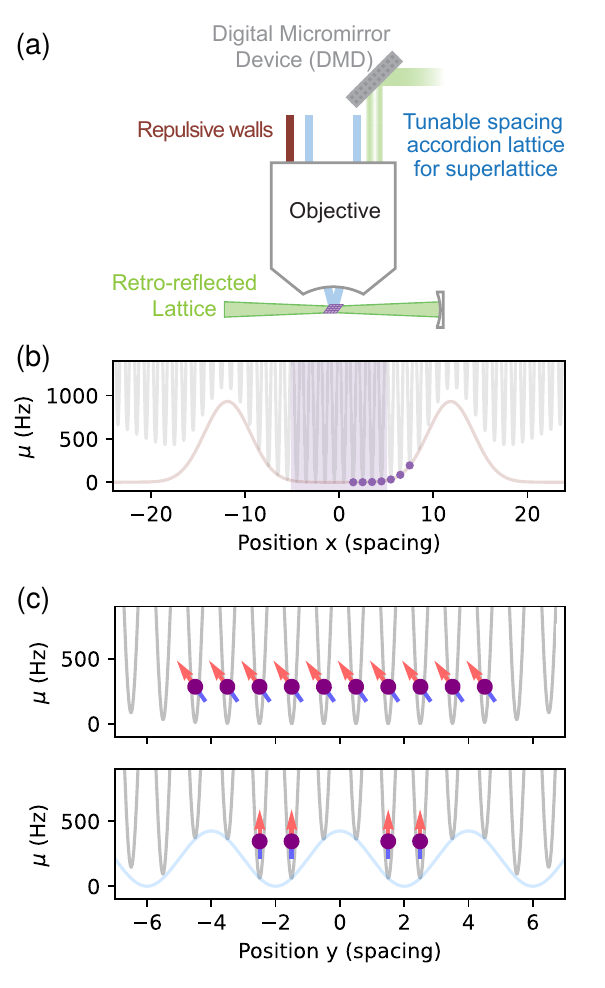}
    \caption{\textbf{Chemical potential.} \textbf{a}, Our microscope allows three types of potential projections: we project superlattices using the blue accordion lattice, project arbitrary potentials using DMD, and project relatively sharp repulsive walls with blue-detuned elliptical beams. \textbf{b}, A schematic chemical potential landscape along the 1D chain for a 10-site system. With DMD to compensate for harmonic confinement and a pair of repulsive walls shown in chestnut color, the central 10 sites (shaded in purple) give us a well-defined system. The rising wall marked in purple dots follows the chemical potential listed in the text. \textbf{c}, To increase the data acquisition rate, we parallel simulate multiple copies of chains. In the top subfigure, we raise the lattice deep to reduce interchain tunneling and tilt the magnetic field to minimize interchain dipole-dipole interactions. In the bottom subfigure, we use an additional superlattice (light blue) to create isolated pairs of chains.
    }
    \label{fig: chemical_potential}
\end{figure}

\subsubsection*{Potential shaping and boundary conditions}

Various optical potentials are projected through the objective, as shown in Fig.~\ref{fig: chemical_potential}a. We use a 532~nm Digital Micromirror Device (DMD) projected through the objective to compensate for the harmonic confinement due to the red detuned optical lattices. Moreover, we use it to introduce programmable chemical potential disorder (Fig.~\ref{fig: CSPT_average}). Due to a custom through-hole in our objective, our projection resolution is limited. The disorder patterns are generated by summing up 50 sinusoidal waves with random amplitudes, phases, and spatial frequencies. They are sampled from uniform distributions: amplitudes from 0Hz to 50Hz, frequencies from 0 to 1/(4 sites), and phases from 0 to $2\pi$. The resulting disorder pattern has a finite correlation length, which can be quantified as 1.1 sites by fitting $\langle\mu_i \mu_{i+d}\rangle$ with an exponential function. The spatial-frequency domain spectrum is similar to white noise with a rounded cutoff.

To realize finite-length chains with relatively sharp edges while minimizing the disorder introduced in the center of the system, we project blue-detuned elliptical beams through the objective to serve as repulsive walls \cite{Gaunt2013}. The beams are roughly 400~MHz detuned relative to a transition with an 8~kHz natural linewidth at 841~nm. When the AC Stark shift is a few hundred Hertz, we estimate a few tens of milliHertz off-resonance scattering rate, which is comparable to the off-resonance scattering rate of the 532~nm optical lattice. Detuning further can improve the off-resonance scattering rate, up to the limitation of the amplified spontaneous emission background of the laser. Boundary conditions play an important role in our study of the Haldane insulator-like state with our finite-size system. Using numerical simulations, we found that the soft rising of the chemical potentials at the edges helps enhance the Haldane string order parameter (Eq.~\eqref{eq: SOP}). As shown by the purple dots in Fig.~\ref{fig: chemical_potential}b, the chemical potentials near the edges are estimated to be [0,1,3,10,32,85,191]~Hz (in comparison $U\sim V=24$~Hz).

\subsubsection*{Experimental steps}

We start with a Bose-Einstein Condensate of $^{164}$Er created within 2 seconds (the reduced abundance and lower scattering length slowed down our experiment compared to~\cite{Phelps2020}). The BEC is then compressed into a 1D vertical lattice and evaporated further to the desired atom number in the region of interest defined by repulsive walls. The 2D lattices are then ramped up adiabatically to the desired tunneling $t$. We begin with an 80-ms exponential ramp with a time constant $\tau = 40$ ms, reaching a lattice depth of roughly three recoil energies. This is followed by a linear ramp of the lattice depth to the target tunneling strength, lasting 120 ms for the CSPT measurements and 200 ms for the MI-HI measurements. Meanwhile, we ramp up the staggered chemical potential $\mu_S$ using superlattices projected through our objective. This allows us to ramp from a BEC (superfluid) to different phases that we study. For Fig.3, after preparing the CSPT-1, we ramp the $\mu_S$ linearly in 100ms, while ramping the perpendicular lattice depth (generating the $t_\perp$) in a parabola.

\subsubsection*{Parallel data acquisition}

To accelerate data acquisition for all measurements except Fig.3, we simulate ~10 decoupled 1D chains in parallel per experiment (Fig.~\ref{fig: chemical_potential}c top). Chains are isolated by minimizing inter-chain tunneling ---ramping lattice depths to reduce tunneling to over an order of magnitude below the intra-chain tunneling, which is itself the smallest term in Hamiltonian (1). For Fig.3, where two coupled chains are simulated, we introduce an additional superlattice to insert two empty rows between each pair (Fig.~\ref{fig: chemical_potential}c bottom). These empty rows have chemical potentials 300~Hz higher, suppressing leakage and allowing inter-chain tunneling up to 100~Hz. We run 2 coupled chains in parallel per experiment.

\subsubsection*{Equivalence between unity-filling Hubbard CSPT, half-filling Hubbard CSPT, and spin-half XXZ CSPTs}

\begin{figure*}
    \centering
    \includegraphics[width=\textwidth]{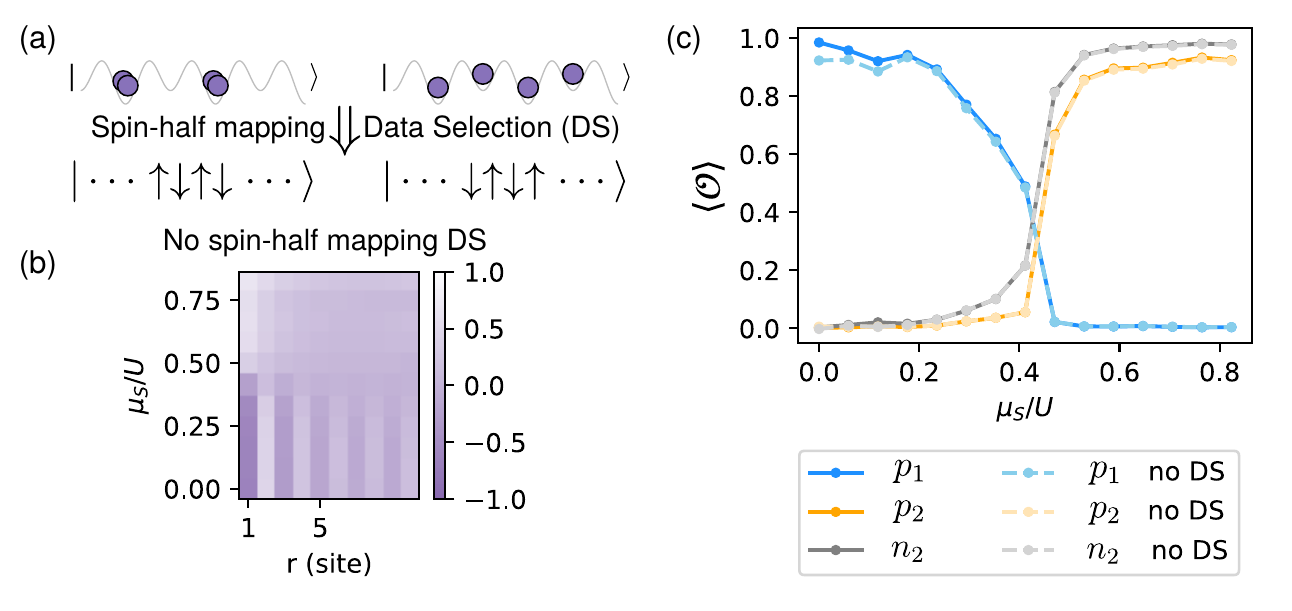}
    \caption{\textbf{Spin-half mapping Data Selection (DS).} \textbf{a}, We argue for the validity of spin-half mapping DS in the main text to reduce the noise. \textbf{b}, We present the data shown in the main text Fig.2b without applying the spin-half mapping DS. The signal of CSPTs is still there - on the bottom half, we observe staggered values, while on the top half, we observe roughly uniform values. \textbf{c}, In the simulated results without experimental imperfections like readout infidelity, the order parameters calculated with and without the spin-half mapping DS overlap with each other near the quantum critical point.
    }
    \label{fig: spin_half_DS_extended_data}
\end{figure*}

In the limit of $t, V\ll U, \mu$, we can map the unity-filling Hubbard model with the staggered chemical potential to an effective spin-half XXZ model with the Hamiltonian

\begin{equation}
\label{eq: XXZ_Hamiltonian}
\begin{split}
H=-\frac{t}{\sqrt{2}}\sum_{\langle i,j \rangle}(X_iX_j+Y_iY_j)+\frac{V}{4}\sum_{\langle i,j \rangle}Z_iZ_j-\\\frac{1}{2}(\frac{U}{2}-\mu_S-V)\sum_i(-1)^iZ_i.
\end{split}
\end{equation}

By mapping to this effective model (Fig.~\ref{fig: spin_half_DS_extended_data}a), we can perform Data Selection (DS) to enhance the signal for the unity-filling CSPT. On sites with attractive staggered potential, the data-selection criterion is one or two bosons; on sites with repulsive staggered potential, the criterion is zero or one boson. This DS effectively reduces the noise coming from doublon-to-empty losses in our system, assisting us in probing $\mathcal{P}_{x_0}(r)$ at a larger distance $r$. To demonstrate that the spin-half mapping is a valid DS criterion, we present two additional data sets. We first present the experimental data without this DS to demonstrate that the signal for the CSPTs is still present, although the measurement noise due to experimental imperfections is worse (Fig.~\ref{fig: spin_half_DS_extended_data}b). Next, we show the DMRG simulated results without accounting for any experimental imperfections (Fig.~\ref{fig: spin_half_DS_extended_data}a). We see no significant difference between before and after the DS close to the quantum critical point.

The mapping can also be extended to the half-filling Hubbard model, which we use to study coupled CSPT chains for enhanced energy gap contrast at finite sizes.

\subsubsection*{Data selection (DS)}

\begin{table*}[ht]
\begin{ruledtabular}
\begin{tabular}{p{0.04\linewidth} | p{0.09\linewidth} | p{0.1\linewidth} | p{0.09\linewidth} | p{0.08\linewidth} | p{0.12\linewidth} | p{0.08\linewidth} | p{0.09\linewidth} | p{0.09\linewidth} | p{0.09\linewidth}}
    Fig & System size (sites) & Atom number required & Atom number DS & Photon count DS & Spin-half mapping DS & Overall DS & Smallest sample size & Average sample size & Total sample size \\
    \hline
    2 & 16 by 1 & 16 & 9.0\% & 41.7\% & 36.2\% & 1.36\% & 197 & 660 & 7261 \\
    3 & 10 by 2 & 10 & 13.1\% & 86.8\% & Not needed & 11.4\% & 250 & 331 & 2975 \\
    4\textbf{a} & 10 by 1 & 10 & 8.1\% & 62.7\% & 53.3\% & 2.72\% & 74 & 204 & 1429 \\
    5 & 10 by 1 & 10 & 8.6\% & 47.4\% & Not applicable & 4.06\% & 99 & 130 & 649 
\end{tabular}
\end{ruledtabular}
\caption{\label{tab:data_size} We apply two or three steps of Data Selection (DS) and show the DS rate. We perform DS based on the total atom number in the system to reduce the effect of wrong initial state preparation or losses during the ramp or imaging processes. In the unity-filling soft-core cases, we ignore the chains containing three or more particles per site, and in the half-filling hard-core case, we ignore the chain pairs containing two or more particles per site. Moreover, we perform DS based on the photon count per site, where the DS rate is lower with a high presence of doublons or triplons or with a large system size. In some cases, we also perform DS based on mapping to the spin-half model. In Fig.3 data, since we map to the half-filling staggered dipolar Bose-Hubbard model, the spin-half mapping DS is unnecessary. In Fig.5 data, while studying the HI, since we no longer project the staggered chemical potential, the spin-half mapping is not applicable. The final number of chains (for Fig.3, the number of double chains) after DS is shown. In each experimental shot, we realize multiple chains in parallel and perform DS on each chain (or double-chain).}
\end{table*}

After taking experimental snapshots, we process the raw photon counts and digitize the filling per site. We throw away the chain(s) when the photon count of a site lies too close to the digitization threshold. We also sum up the total atom number in the chain(s) and throw away the data where we do not have the required filling. When studying the tunnel-coupled pair of chains in Fig.~\ref{fig: Coupled_CSPT} mapped to half-filling, we sum up the total atom number in two chains and require that the number of bosons equals half of the number of sites in the two-chain system. Specifically, in the case of unity-filling CSPTs, we apply one more DS by mapping the system to a spin-half Hamiltonian and throwing away the data that is not within the spin-half basis. The DS rates and final data set sizes are presented in Table~\ref{tab:data_size}. We emphasize that the DS only serves to enhance the signal, and the phases we explore are not a product of post-selection.

\subsubsection*{Mapping for coupled-chain simulations}

To perform a quantum simulation of the $H_\perp$ coupling in main text Eq.~\eqref{eq: coupling} in an atom-number-conserving system, we map the unity-filling CSPTs to the half-filling ones and perform a particle-hole transformation on the bottom layer, treating the \textit{physical bosons} present in the bottom layer as holes (main text Fig.~\ref{fig: Coupled_CSPT}a bottom).
In this representation, the pairing term ($\Delta_\perp$) becomes an inter-layer hopping term ($t_\perp$), which can be implemented experimentally. The staggered potential is created by a diagonal superlattice, resulting in a checkerboard pattern.

\subsubsection*{Two-body and three-body inelastic losses at a single site}

Inelastic losses can reduce the ground state fidelity and hurt the DS rate of our quantum simulator. We noticed that the inelastic loss rate is the best when we are far away from the Feshbach resonances~\cite{Chin2010, Frisch2014}. $^{164}$Er is used because the effective S-wave scattering length can be tuned to the required values at low fields far from Feshbach resonances~\cite{Patscheider2022}. Using parity-projection-free site-resolved imaging~\cite{Su2024}, in our main 532~nm wavelength lattice, we measure the lifetime of the doublons to be more than 1 second and the lifetime of the triplons to be more than 150 milliseconds. As a comparison, our typical ramp duration is 150 milliseconds, and doublon and triplon occupancy is not significant for most of the ramp durations. The expansion of the 488~nm wavelength accordion lattice takes another 80 milliseconds, and the lifetime of doublons is measured to be highly dependent on the Wannier function size. Due to these infidelities, the observables in Fig.2c do not reach zero exactly.

\subsection*{Numerical simulation}

\subsubsection*{Ramp simulation}
\begin{figure}
    \centering
    \includegraphics[width=0.48\textwidth]{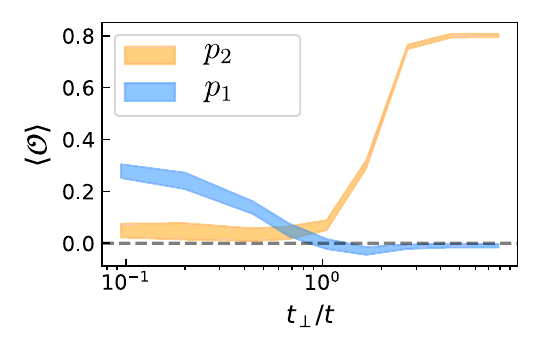}
    \caption{\textbf{Simulated order parameters without any decoherence.} In the main text Fig.3, we demonstrated the simulated order parameters after taking into account decoherences due to the diagonal superlattice. Here we show the simulated results without any decoherence, where a relatively sharp crossover of the order parameter can be seen.
    }
    \label{fig: Coupled_CSPT_no_decoherence}
\end{figure}

\paragraph*{CSPT simulations} For the theory calculations shown as shades in the figures, we simulate the ramp from the superfluid to the final state in our finite-size system using the tensor network algorithms on the ITensor environment~\cite{itensor}. Ground state calculations are performed with density matrix renormalization group (DMRG) with a maximum bond dimension of $\chi_m=300$ and a truncation error cutoff of $\epsilon \approx 10^{-12}$ for matrix product state (MPS) compression. The exact ramp sequences of the experiment are implemented via Trotter decomposition and time-evolving block decimation (TEBD), where each step $\chi_m=300$ and $\epsilon \approx 10^{-10}$ are set with a time step of $dt=0.2$ms. All simulations set a maximum boson number of 3 per site. We also take into account the particle fluctuations in the experiment, i.e.,~configurations slightly away from the unity filling at chain size $L$, such as $N=L, L\pm1, L\pm2$ particles. It is crucial to model these particle fluctuations to capture the experimental conditions, and consequently, we find that such particle fluctuations slightly suppress the CSPT order parameters throughout the phase diagram, even after Data Selection (DS) based on total particle number, because of losses and other imaging infidelities. For the simulation of the HI-MI transition, we take into account the boundary conditions of our Gaussian-shaped walls.

The experiment reported in Fig.~\ref{fig: Coupled_CSPT} utilizes a diagonal superlattice to project the chemical potential of the required checkerboard pattern. We experimentally observe that the laser utilized for this chemical potential causes significant heating compared to the rest of the experiments, where we used non-diagonal superlattices. This heating causes non-negligible decoherence in the experiment, which has to be modeled to capture the experimental results. Without taking into account the decoherence, we obtain simulated results shown in Fig.~\ref{fig: Coupled_CSPT_no_decoherence}. Here we choose to simulate the dissipative time evolution with a classical white noise, which emulates a Lindbladian master equation with a jump operator $\hat n_i$~\cite{Kampen1992, Seif2022}. For a faithful representation of dissipative dynamics with correlated classical white noise, we introduce a time-varying chemical potential to the Hamiltonian, Eq.~\eqref{eq: Hamiltonian}, i.e.,~$H_{\rm wn}=\sum_i \mu_i(t) \hat  n_i$. Here $\mu_i(t)$ is distributed normally with $\braket{\mu_i(t)}=0$ and $\braket{\mu_i(t)\mu_j(t')}=\gamma_{ij}(t) \delta(t-t')$ where $\Gamma(t)=\left[\gamma_{ij}(t)\right]=\sigma(t)\mathcal{I}$ is the covariance matrix for classical white noise at time $t$, i.e.,~$\mu_i(t)$ for different sites $i$ are sampled from a Gaussian distribution with a mean of $0$ and a standard deviation of $\sigma(t)$. Because the diagonal superlattice also governs the staggering potential $\mu_s(t)$, we modulate $\sigma(t)$ with $\mu_s(t)$, i.e.,~$\sigma(t)=\sigma_0 |\mu_s(t)|/\textrm{max}(|\mu_s(t)|)$. In other words, the most heating occurs during the time evolution when we are deep in the CSPTs, e.g.,~$\mu_s \approx \pm 100$~Hz, and it becomes negligible around the critical point, e.g.,~$\mu_s=0$. Under these conditions, we find that $\sigma_0=20$~Hz successfully models the experimental results for all $t_\perp/t$ in Fig.~\ref{fig: Coupled_CSPT}b.

Let us note that, although the decoherence is non-negligible, by comparing the experimental data to the ED results of a smaller system size, we estimate that the temperature is less than half of the energy gap, allowing us to probe ground-state physics.

\paragraph*{MI-HI simulations} For the numerics performed to simulate the ramp from the MI to the HI, we simulated 16 sites with the edge chemical potentials described in the previous section.
The numerical simulations were performed using the Tensor Network Python (TeNPy)~\cite{Hauschild18}. 
We found that a bond dimension of $\chi_{\text{max}} = 182$ was sufficient to guarantee convergence for the systems and parameters considered.
The simulations were performed by initializing the simulations the ground state of a superfluid with different total atom numbers found using DMRG, and the dynamical ramps were simulated using the matrix product operator-based approach developed
in Ref.~\cite{Zaletel2015} with a trotter step of $dt = 0.1 \text{ ms}$.

\subsubsection*{Ground state simulations}

\paragraph*{Local observables stay non-sharp at large system sizes}

\begin{figure}
    \centering
    \includegraphics[width=0.48\textwidth]{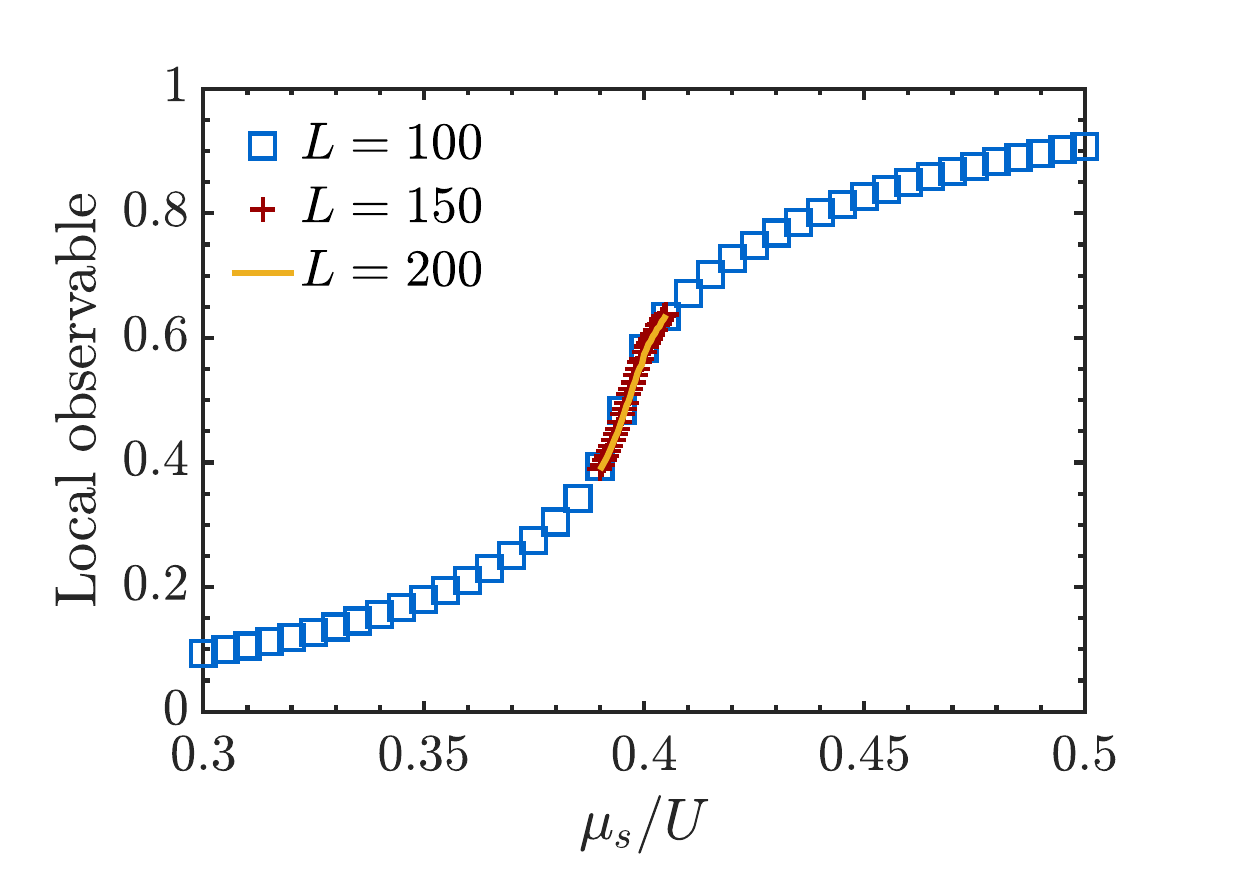}
    \caption{\textbf{Numerical simulation of larger system sizes shows that local observables stay non-sharp.} To demonstrate the non-sharpness of Fig. 2 local observable is not due to finite system size, we perform DMRG simulations to probe larger sizes and plot $n_2$. U/t=20, V/t=2.5, atom number per site truncation is 4, dipolar tails are on according to the experimental values.
    }
    \label{fig: large_size_local_observable}
\end{figure}

In the main text Fig.2, we presented that the local observable $n_2$ varies smoothly and argued that this cannot serve as an order parameter because it is not zero out of the CSPT-2 phase. Our experiment is at a finite system size of $L=16$ sites. To argue that such a behavior is not due to finite-size effects, we use DMRG to simulate the system at $L=100, 150$ and $200$ sites and observe the same smoothness, as shown in Fig.~\ref{fig: large_size_local_observable}.

\paragraph*{Coupling of multiple $\mathbb{Z}_2$ CSPTs}

\begin{figure}
    \centering
    \includegraphics[width=0.48\textwidth]{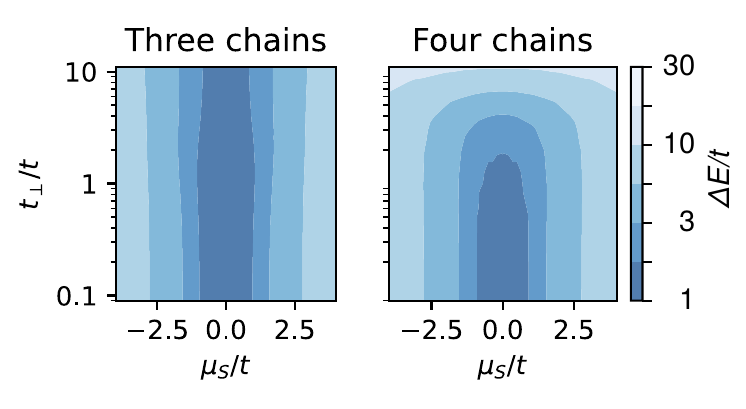}
    \caption{\textbf{Energy gap simulation for coupling more than two CSPT chains.} Coupling three (odd number) chains revives the quantum phase transition between the two phases, as can be seen by the parallel contour lines. Coupling four (even number) chains gives an energy gap similar to coupling two chains, where the gap closing is gone at high $t_\perp$ values.
    }
    \label{fig: three_or_more_coupled_chains}
\end{figure}

In the main text, we presented that decoupled chains have two topologically distinct CSPTs, but two tunnel-coupled chains trivialize the topology. Using Exact Diagonalization (ED) (\textsc{QuSpin}~\cite{QuSpin}) with three by eight sites and four by six sites, we further numerically simulate higher numbers of coupled chains to demonstrate the $\mathbb{Z}_2$ properties as shown in Fig.~\ref{fig: three_or_more_coupled_chains}.

\paragraph*{Interpolation between CSPT and MI-HI quantum phase transitions}

\begin{figure*}
    \centering
    \includegraphics[width=\textwidth]{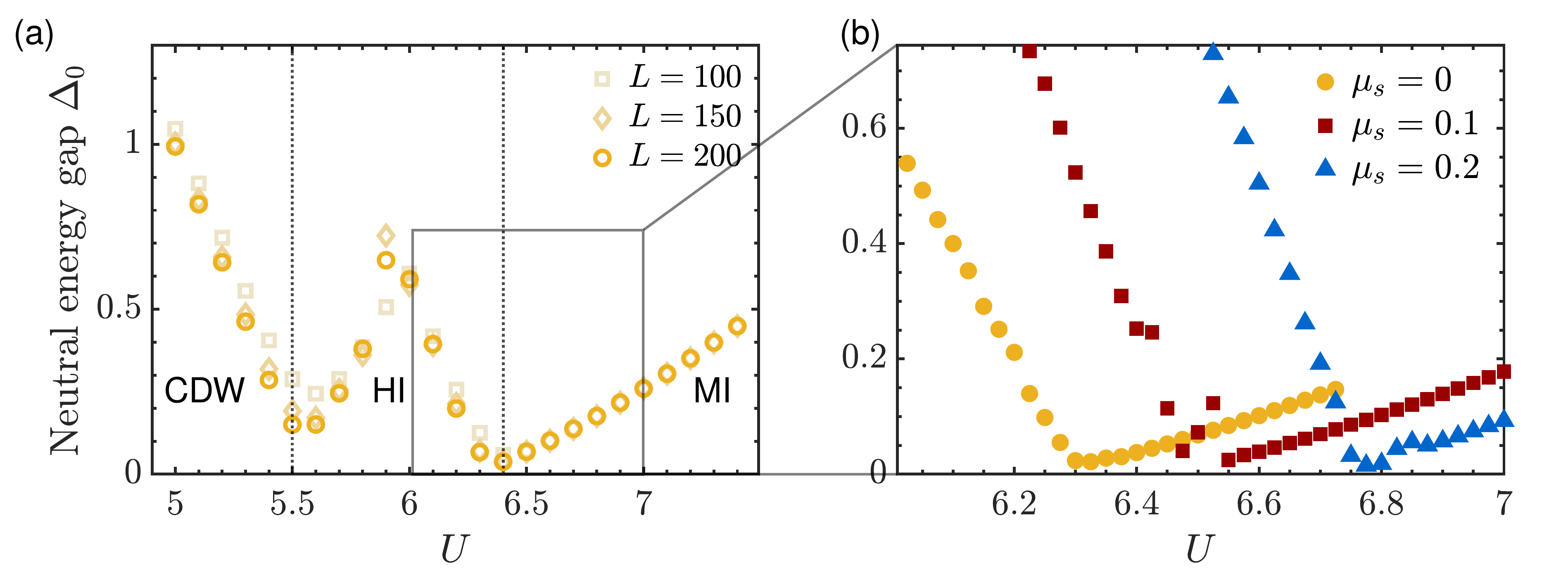}
    \caption{\textbf{Simulated energy gap indicates the connection between MI-HI transition and the CSPT transition.} \textbf{a}, The energy gap of the MI-HI transition at $\mu_S=0$ for different system sizes ($L$). \textbf{b}, Turning on $\mu_S$ at fixed system size $L=256$ sites, the MI-HI transition interpolates to the CSPT transition. All simulations in this Figure are performed with interactions up to the nearest neighbor with a maximum boson number of 2 per site for fast convergence in DMRG.
    }
    \label{fig: HI_to_CSPT_transition_large_size}
\end{figure*}

We show in Fig.~\ref{fig: HI_to_CSPT_transition_large_size} the energy gap for the neutral excitations, $\Delta_0=E_1-E_0$, where $E_0$ and $E_1$ are the ground and the first excited states at a fixed system size with unit filling. Subfigure a demonstrates the gap closings at two phase transitions between CDW, HI, and MI phases, which become sharper as the system size increases. Subfigure b focuses on the HI-MI transition and shows how turning on the staggered chemical potential $\mu_s$ transforms the MI-HI transition to the CSPT transition.

\subsection*{Theoretical analysis}

In this section, we provide some additional details for the theoretical analysis behind our experimental results.
While the section below leaves the present work self-contained, an extensive and pedagogical theoretical accompaniment is provided in the forthcoming \cite{Sahay2025} manuscript.

\subsubsection*{Crystalline symmetry}
We start by making a brief remark on the nature of crystalline symmetries for finite systems.
Strictly speaking, the system only has a site-centered inversion symmetry for odd-length systems.
Nevertheless, for the bulk phase of matter properties, this distinction between even and odd system sizes should not play a role.

\subsubsection*{CSPT Topological invariant}

To get an intuitive understanding of the topological invariant that distinguishes CSPT-1 and CSPT-2, let us first work with the product states that appear in the atomic limit.
One way to distinguish these states is to recognize that in any large inversion-symmetric region, the parities of bosons in CSPT-1 and 2 are $-1$ (odd) and $+1$ (even) respectively (main text Fig.~\ref{fig: CSPT_ground}a for CSPT-1, top).
As quantum fluctations are introduced and we move away from the atomic limit, the bosons in these states delocalize into dressed quasiparticles (main text Fig.~\ref{fig: CSPT_ground}a, bottom) and the expectation value of the parity in this region is no longer exactly $+1$ or $-1$.
Nevertheless, the \textit{sign} of the expectation value of the parity in these regions is a discrete invariant that can be used to distinguish the two.
As long as the gap stays open and both parity and inversion symmetry are preserved, any change in the parity expectation within the region occurs symmetrically from its edges. As a result, the sign of the bulk parity remains unchanged (main text Fig.~\ref{fig: CSPT_ground}a, bottom).
The technical tools required to derive this invariant are in Ref.~\cite{Fuji2015} and follows closely the derivation of the mixed state invariant provided later in the SM.

\subsubsection*{Absence of local order parameters in SPTs}
The Landau paradigm characterizes phase transitions by local order parameters that signal spontaneous symmetry breaking~\cite{Landau1937, Beekman2019}. In sharp contrast, the transition we study lies outside this framework: it separates two symmetry-protected topological (SPT) phases that are indistinguishable by any local observable. While local observables such as the average site occupation respond to the staggered chemical potential, they cannot serve as order parameters for distinguishing the two crystalline SPTs. This is because the Hamiltonian already breaks translational symmetry explicitly, so density modulations are always present—even in trivial phases—and they vary smoothly across the transition. To serve as an order parameter, the observable has to be zero outside the phase and non-zero inside the phase. In contrast, the topological invariant distinguishing the phases (e.g., the sign of the inversion-symmetric parity string) changes singularly and is protected by symmetry and a finite gap~\cite{Fuji2015, Pollmann2012}.

More formally, we note that any two symmetric, short-range entangled ground states cannot be distinguished by the expectation value of a local observable, since symmetric, finite-depth local unitary circuits can map local operators to other local operators without altering the symmetry constraints~\cite{Chen2010, Chen2011}. Consequently, the expectation value of any symmetric local observable can always be smoothly deformed between different SPTs, and cannot serve as a sharp order parameter.

\subsubsection*{Theoretical Framework for Mixed State Order}

The experimental realization of mixed state order can be formalized as follows. We draw disorder realizations $\mathcal{D}$ from a distribution $p_{\mathcal{D}}$ such that disorder patterns related by site-centered inversion are equally likely.
Then the ground state for each disorder realization $\ket{\psi_{\mathcal{D}}}$ is probed by the following
\begin{equation}
    \mathbb{E}[\mathcal{P}_{x_0}(r)] = \sum_{\mathcal{D}} p_{\mathcal{D}} \bra{\psi_{\mathcal{D}}} \hat{\mathcal{P}}_{x_0}(r)\ket{\psi_{\mathcal{D}}}.
\end{equation}
Crucially, probing the above is equivalent to experimentally probing the expected value of the parity string operator in the mixed state $\rho = \sum_{\mathcal{D}} p_{\mathcal{D}} \ket{\psi_{\mathcal{D}}} \bra{\psi_{\mathcal{D}}}$ with $\mathbb{E}[\mathcal{P}_{x_0}(r)] = \text{Tr}\left( \rho\,  \hat{\mathcal{P}}_{x_0}(r)\right)$.

Now, we prove that the crystalline SPT phases we study persist provided that the site-centered inversion symmetry of the chain is preserved on average.
Before delving into the proof, let us review some of the foundational concepts in the many-body physics of mixed states that will be necessary for our analysis.

\textit{Short-Range Entanglement and Symmetries in Mixed States.} When generalizing the concept of an SPT from pure states to mixed states, it is essential to understand both the entanglement structure of mixed states as well as their symmetries.

A central property of pure state SPTs is that they are short-range entangled (SRE)---i.e. they can be prepared from a product state by finite time evolution with a local Hamiltonian.
To discuss SPTs in the mixed state context, we need to find an appropriate mixed state generalization.
While multiple notions have appeared in literature\cite{Ma2023, Chen2024}, in our case, if $\rho$ is the density matrix of a mixed state defined on a Hilbert space $\mathcal{H}_A$, we say it is SRE if there exists a SRE pure state $\ket{\psi_{\text{AB}}} \in \mathcal{H}_A \otimes \mathcal{H}_B$ such that $\rho = \text{Tr}_{B}(\ket{\psi_{\text{AB}}} \bra{\psi_{\text{AB}}})$.
In other words, $\rho$ is SRE if it has an SRE purification.

Having defined a notion of SRE for mixed states, we now turn to understanding their symmetries.
For mixed states, there are conceptually two different types of symmetries.
In particular, if we suppose a mixed state $\rho_A$ arises from a pure state in a larger system defined on a Hilbert space $\mathcal{H}_{A} \otimes \mathcal{H}_{B}$, the larger system can either have a symmetry that acts purely on $\mathcal{H}_A$ or collectively on $\mathcal{H}_A \otimes \mathcal{H}_B$.
Equivalently, there can exist conserved quantities that reside purely in $A$ or are exchanged between $A$ and $B$.
In the first case, we say that the system has a \textit{strong}  (or exact) symmetry, where the symmetry generated by an operator $U_A \otimes \mathds{1}_{B}$ and leaves the mixed state invariant under right or left multiplication by $U_A$, i.e.~ $U_A \rho_A = e^{i \theta} \rho_A$.
In the second case of \textit{weak} (or average) symmetries, the symmetries are instead generated by the operator $U_A \otimes U_B$ and the density matrix is only invariant under conjugation, i.e.~  $\rho = U_A \rho U_A^{\dagger}$.

\textit{SRE and Symmetric Purifications of the Staggered Dipolar Bose-Hubbard Model with Disorder.} We now wish to show that the ensemble of ground states of the staggered dipolar Bose-Hubbard (SDBH) model with weak disorder is an SRE mixed state with a strong $\mathbb{Z}_2$ parity symmetry and an average site-centered inversion symmetry.
To do so, we show explicitly that the ensemble of ground states of the staggered dipolar Bose-Hubbard model with disorder has an SRE purification that is parity symmetric and inversion symmetric.
To do so, let us suppose that the distribution of the disorder on each site is uncorrelated\footnote{This assumption is not necessary and is made for simplicity. In the experiment, it is not exactly realized due to the finite resolution of potential projection} such that the probability of each disorder realization of the chemical potential $\mathcal{D} = \{\mu_x\}$ is $p(\mathcal{D}) = \prod_{x} p_x(\mu_x)$.
Then, we find the purification as follows.

Let $H_A$ be the SDHB model (Eq.~\eqref{eq: Hamiltonian}) without any disorder in the regime where $U$ is large and $\mu_{\text{S}}$ is tuned far away from criticality.
Furthermore, let $\mathcal{H}_B$ be the Hilbert space of real-valued states $\ket{\mu_x \in \mathbb{R}}$ defined at each lattice site $x$, which satisfy $\langle \mu_x|\mu'_x\rangle =  \delta(\mu_x - \mu'_x)$ and
whose Hamiltonian is: 
\begin{equation}
    H_{B} = -\Delta \sum_{x} \ket{p_x} \bra{p_x} \qquad \ket{p_x} = \int d\mu_x \sqrt{p_x(\mu_x)} \ket{\mu_x}  
\end{equation}
where $\Delta$ will be taken to infinity.
Lastly, take the coupling between the $A$ and $B$ system to be: 
\begin{equation}
    H_{\text{AB}} = - \sum_{x} n_x \hat{\mu}_x
\end{equation}
where $\hat{\mu}_x \ket{\mu_x} = \mu_x \ket{\mu_x}$. 
The ground state of the Hamiltonian $H = H_{\text{A}} + H_{\text{B}} + H_{\text{AB}}$ (in the $\Delta \to \infty$ limit) can be easily expressed in terms of the ground states of the disordered SDBH model. 
In particular, note that $H_B$ projects the $\mu$ degrees of freedom into the $\ket{p_x}$ state, and the $H_{\text{AB}}$ term makes the $A$ system see the $B$ system as an effective chemical potential.
Consequently,
\begin{equation}
    \ket{\psi_{\text{AB}}} = \sum_{\{\mu_x\}} \underbrace{\left(\prod_{x} \sqrt{p(\mu_x)} \right)}_{p_{\mathcal{D}}} \ket{\psi[\{\mu_x\}]}_{\text{A}} \otimes \ket{\{\mu_x\}}_{\text{B}}
\end{equation}
where $\ket{\{\mu_x\}} = \bigotimes_x\ket{\mu_x}$ and $\ket{\psi[\{\mu_x\}]}$ is the ground state of the SDBH model with disorder configuration $\{\mu_x\}$.
Let us note that $\ket{\psi_{\text{AB}}}$ is a purification of the ensemble of ground states of the SDBH model, i.e. 
\begin{equation}
    \rho_A = \sum_{\mathcal{D}} p_{\mathcal{D}} \ket{\psi_{\mathcal{D}}} = \text{Tr}_B\left(\ket{\psi_{\text{AB}}}\bra{\psi_{\text{AB}}}\right).
\end{equation}

At this point, a few remarks are in order. 
First, let us note that the Hamiltonian $H$ is gapped; this can be seen by noting that $H_B$ is exactly solvable and gapped, $H_A$ is gapped, and provided that the disorder is sufficiently weak, $H_{\text{AB}}$ is a small perturbation that cannot close the gap. 
As a consequence, $\ket{\psi_{\text{AB}}}$ is the unique gapped ground state of a Hamiltonian and can be prepared via quasi-adiabatic continuation from a product state and hence is SRE.
Thus, $\rho$ admits an SRE purification.
In addition note that since $\ket{\psi_{\text{AB}}}$ is symmetric under parity on $A$ and is site-centered inversion symmetric under $\mathcal{I}_A \otimes \mathcal{I}_B$, $\rho$ has a strong $\mathbb{Z}_2$ parity symmetry and an average $\mathcal{I}_A$ symmetry as desired.

\begin{figure*}
    \centering
    \includegraphics[width=\textwidth]{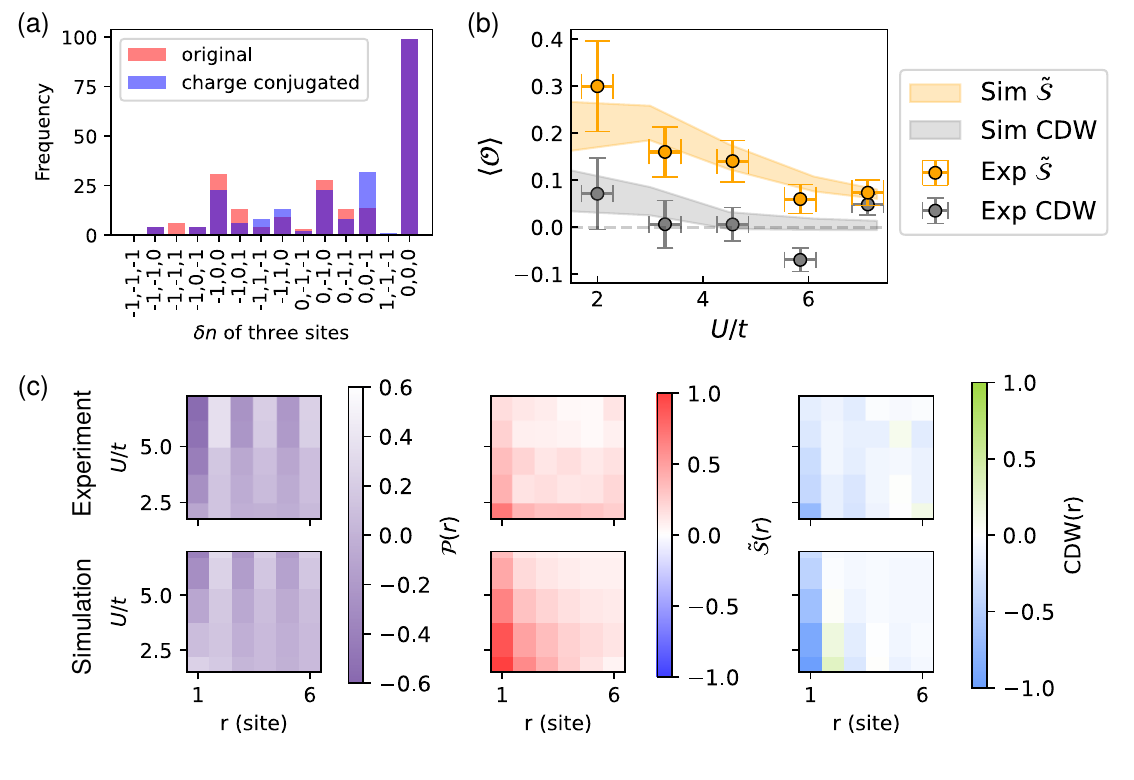}
    \caption{\textbf{Order parameters for the Mott insulator to Haldane insulator quantum phase transition.} \textbf{a}, We randomly pick three sites in the system and plot the histogram of the observed states. The similarity of the original (red) and charge conjugated (blue) counts serves as experimental evidence of the approximate charge conjugation symmetry. \textbf{b}, We compare the Haldane string order parameter and the CDW order parameter after normalization to demonstrate that we are in the Haldane insulator regime. The vertical error bars are obtained from bootstrapping standard error; the horizontal error bars are estimated standard error. \textbf{c}, Observables plotted with respect to $r$.
    }
    \label{fig: HI_2D_OP}
\end{figure*}

\textit{Average SPT Invariants from Exact Parity and Average Inversion Symmetry.} We now prove that the parity string order parameter measures a topological invariant of this ensemble that cannot be changed under a finite-depth quantum channel
To do so, let us suppose that $\rho$ is an SRE mixed state that is invariant under a strong $\mathbb{Z}_2$ symmetry $\mathcal{P}_A$ and a weak site-centered inversion symmetry $\mathcal{I}_A$.
Then, $\rho$ has an SRE purification $\ket{\psi_{\text{AB}}}$ symmetric under $\mathcal{P}_A \otimes \mathds{1}$ and $\mathcal{I}_A \otimes \mathcal{I}_B$.
Therefore, we have that for $a, b$ separated at a distance much longer than the correlation length $\xi$, $\mathbb{Z}_2$-symmetry string $\mathcal{P}_{a, b}$ fractionalizes on the state as:
\begin{equation}
    \mathcal{P}_{a, b} \otimes \mathds{1} \ket{\psi_{\text{AB}}} = q^L_a q^R_b \ket{\psi_{\text{AB}}}
\end{equation}
where $q^L_a$ and $q^R_b$ are unitary operators that are localized at the endpoints of $a$ and $b$ respectively (but potentially supported on both the $A$ and $B$ subsystem).
Let us note that there is an ambiguity in the definition of $q^L_a$ and $q^R_b$ in that we can take $q_{a}^L \to e^{i \alpha} q_a^{L}$ and $q_{b}^{R} \to e^{-i \alpha} q_a^{L}$ without changing the above relation (where $\alpha \in U(1)$ is a phase independent of $a, b$).
We can fix this freedom up to a minus sign by demanding that $q_a^L$ and $q_b^R$ square to one (i.e., form a representation of $\mathbb{Z}_2$), which can be done since $\mathcal{P}_{a, b}$ squares to one.

With this in mind, let $[a, b]$ be an odd inversion symmetric region of the chain; without loss of generality, we may take $b = -a$.
Then under the inversion symmetry $\mathcal{I}_0 = \mathcal{I}_{A, 0} \otimes \mathcal{I}_{B, 0}$ around the origin, we have that:
\begin{align}
    \mathcal{I}_{0} \mathcal{P}_{a, -a} \otimes \mathds{1} \ket{\psi_{\text{AB}}} &= (\mathcal{I}_0 q_{a}^L \mathcal{I}_0)(\mathcal{I}_0 q_{-a}^R \mathcal{I}_0) \ket{\psi_{\text{AB}}} \nonumber\\
    &= q_{a}^L q_{-a}^{R}\ket{\psi_{\text{AB}}} = \mathcal{P}_{a,-a} \ket{\psi_{\text{AB}}}
\end{align}
The above implies that: 
\begin{equation}
     (\mathcal{I}_0 q_{a}^L \mathcal{I}_0) = e^{i \Theta_a} q_{-a}^{R} \qquad (\mathcal{I}_0 q_{-a}^R \mathcal{I}_0) = e^{-i \Theta_a} q_{a}^{L}
\end{equation}
Now, since $q_a^{L/R}$ squares to the identity, we know from the above that $\Theta_a$ can only take on values $0, \pi$.
Since $\Theta_a$ is a discrete number characterizing the state, it cannot be changed continuously with a symmetric quantum channel on $\rho$ (equivalently, an FDLU on $\ket{\psi_{\text{AB}}}$).
We remark that $\Theta_a$ can be extracted as: 
\begin{align}
    \text{Tr}\left(\rho \mathcal{P}_{a, -a} \right) = \langle q_a^L q_{-a}^R \rangle_{\psi_{\text{AB}}} &\approx \langle q^L_a \rangle_{\psi_{\text{AB}}} \langle q^R_{-a} \rangle_{\psi_{\text{AB}}}\\
    &= e^{i \Theta_a} \langle q_a^L \rangle^2_{\psi_{\text{AB}}}
\end{align}
where the second step comes from the clustering of correlations in SRE states, and in the last step we used the fact that $\langle q_{-a}^R \rangle_{\psi_{\text{AB}}} = \langle \mathcal{I}_0 q_{-a}^R \mathcal{I}_0 \rangle_{\psi_{\text{AB}}} = e^{i \Theta_a} \langle q_a^{L} \rangle_{\psi_{\text{AB}}}$.
Consequently, the sign of the parity string for odd-length regions is a topological invariant characterizing the state.
Note that, in principle, for every pair of inversion-related sites, there is a topological invariant $\Theta_a$.
However, in the presence of a translation symmetry (e.g., the two-site translation that arises if site-centered inversion is a symmetry around any site), only two invariants remain \cite{Sahay2025}.

\subsection*{Haldane to Mott insulator phase transition}

In this section of the SM, we provide some additional intuition for the connection betwen the Haldane to Mott insulator transition and the transition between CSPT-1 and CSPT-2.

\begin{figure}
    \centering
    \includegraphics[width=0.48\textwidth]{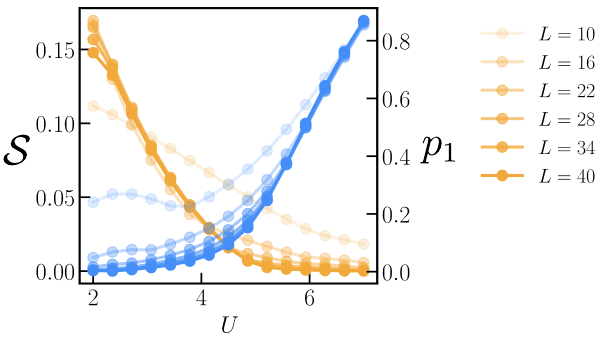}
    \caption{\textbf{Finite-size scaling simulation for MI-HI transition.} At larger system sizes, the order parameters become sharper, indicating a quantum phase transition. $V_\mathrm{nn}/t=3$, $V$ follows the experimental profile that decays as $1/r^{2.7}$. The boson number per site is truncated to three.
    }
    \label{fig: finite_size_scaling}
\end{figure}

\subsubsection*{Connection between the MI-HI transition and the CSPT transition}

Let us recall that, in the absence of the staggered potential ($\mu_{\text{S}} = 0$), the phase diagram of the dipolar Bose-Hubbard model is theoretically predicted to contain at least three insulating phases as a function of the on-site interaction strength $U$ and the characteristic strength of the long-range interactions $V$ (schematically reproduced in main text Fig.~\ref{fig: HI}a)~\cite{Torre2006, Berg2008}. 
Two of these are the Mott insulator (MI) and the charge density wave (CDW), the latter of which spontaneously breaks the translation symmetry $\mathbb{Z}$ of the model at $\mu_{\text{S}} = 0$ down to $2\mathbb{Z}$.
The last phase is the Haldane Insulator (HI)---an entangled SPT phase that is separated from the Mott insulator by a quantum phase transition in the presence of the crystalline \textit{bond-centered} inversion symmetry and parity symmetry of the dipolar Bose-Hubbard model at $\mu_{\text{S}} = 0$~\cite{Pollmann2010}.
In this sense, the HI is a CSPT phase with protected quantum entanglement.

To see how the Haldane to Mott insulator transition is connected to the transition between CSPT-1 and CSPT-2 \cite{Sahay2025}, let us note that under adding a staggered potential, the Mott insulator morphs into CSPT-1 manifestly. 
To gain an intuitive understanding of the relationship between HI and CSPT-2, we use the ``dressed quasi-particle'' picture of the CSPTs introduced in the main text (Section~\ref{sec:CSPT-transitions}).

In particular, the HI can be thought of as a case where bosons in our system are each delocalized across the bonds of the lattice, as shown in the main text Fig.~\ref{fig: HI}a---a configuration that is symmetric under the parity, site- and bond-centered inversion symmetries of the model.
Concretely, a model wavefunction for this phase is given by,
\begin{equation} \label{eq:model-wavefunction}
    \ket{\psi} \propto \bigotimes_{x} \left( b_{x}^{\dagger} + b_{x + 1}^{\dagger} \right) \ket{0},
\end{equation}
which captures the bond delocalization.
The string order parameter in the main text Eq.~\eqref{eq: SOP} is similar to the parity string operator we used for CSPT-1 and CSPT-2; however, now there is an endpoint dressing $\delta \hat n = \hat n-1$. The reason for this is that without the dressing, the expected parity would vanish since the bond-delocalized bosons are equally likely to be inside as outside the region of interest. The dressing accounts for this, and the nonzero value of $\mathcal{S}_{x_0} (r)$ is predicted to be positive (for odd lengths), precisely consistent with the phenomenology of CSPT-2 and revealing their connection.
Note that the endpoint operator $\delta n$ transforms to $-\delta n$ under the approximate charge-conjugation symmetry of the chain---it is precisely the transformation law of this endpoint decoration that serves as a topological invariant of the HI \cite{Pollmann2012, Schuch2011, Chen2011}.

\subsubsection*{Approximate charge conjugation symmetry}
To demonstrate the approximate charge conjugation symmetry, we analyze the central 3 sites of the system and plot a histogram of the distribution. The distribution is roughly symmetric with respect to the charge conjugation center, demonstrating the approximate symmetry.
This justifies the use of the string order parameter $\mathcal{S}$ to identify the HI state and further justifies the vanishing of the effective string observable $p_1$ ($p_2$) in CSPT-$2$ (CSPT-$1$).

\subsubsection*{Extended data}
We present the observable in 2D color plots for the Mott insulator to Haldane insulator quantum phase transition in our finite-size experimental system in Fig.~\ref{fig: HI_2D_OP}. To distinguish the Haldane insulator phase from the nearby Charge Density Wave phase, we also plot the normalized two-point density-density correlators $\mathrm{CDW}(r)=\eta\langle \delta n_i \delta n_{i+r} \rangle$. We compare the CDW order parameter with the Haldane string order parameter after normalizing both of them by $\eta$ defined in the main text.

The sharpness of the quantum phase transition probed between the Haldane insulator and the Mott insulator~\cite{Fraxanet2022} is limited by the finite size (10 sites) practically accessible in our experiment. To numerically clarify the quantum phase transition between the Mott insulator and the Haldane insulator, we perform finite-size scaling using DMRG (Fig.~\ref{fig: finite_size_scaling}). We note that our system size is smaller than the simulated correlation length in the thermodynamic limit using iDMRG. The disorder and decoherence in our experimental system may reduce the correlation length, but we still want to note that the states we probe in this experiment may be in the quantum critical regime and strictly speaking not insulating. Moreover, the walls that set the system size are not infinitely sharp, and certain leakage of atoms outside of the defined system may occur, especially when $U$ is large.

Future advancement in quantum gas microscopes may allow for the measurement of the entanglement spectrum for a large enough system size, revealing signatures of the edge modes of the Haldane insulator phases in the dipolar Bose-Hubbard model~\cite{Deng2011, Tran2023}.

\section*{Additional Information}

\paragraph*{Data and code availability}
The experimental data and analysis code supporting this study's findings are available from the corresponding authors upon request.

\paragraph*{Acknowledgements}
The experimental team is grateful for the contributions to building the experiment from A. Hebert, A. Krahn, G. Phelps, R. Groth, S. F. Ozturk, S. Ebadi, S. Dickerson, F. Ferlaino, V. Singh, and V. Kaxiras. We acknowledge fruitful discussions with Hans Peter B\"uchler, Daniel Mark, Lode Pollet and Yuxin Wang, and we thank Ruochen Ma for insightful comments on the manuscript. 
We are supported by U.S. Department of Energy Quantum Systems Accelerator DE-AC02-05CH11231, National Science Foundation Center for Ultracold Atoms PHY-1734011, Army Research Office Defense University Research Instrumentation Program W911NF2010104, Office of Naval Research Vannevar Bush Faculty Fellowship N00014-18-1-2863, Gordon and Betty Moore Foundation Grant GBMF11521, and Defense Advanced Research Projects Agency Optimization with Noisy Intermediate-Scale Quantum devices W911NF-20-1-0021. A.D. acknowledges support from the NSF Graduate Research Fellowship Program (grant DGE2140743). R.S. acknowledges support from the Department of Energy Computational Science Graduate Fellowship (CSGF) under Award Number DE-SC0022158. C.D. was supported with the ITAMP grant No. 2116679.

\paragraph*{Author contributions}
L.S., M.S., A.D., and O.M. contributed to building the experiment setup. L.S. acquired and analyzed the data in the experiment. C.D., R.S., and L.S. performed numerical simulations. R.S., C.D., and R.V. developed the theoretical framework, and R.S. and R.V. developed mathematical proofs. L.S., R.S., C.D., M.S., R.V., and M.G. contributed to writing the manuscript. All authors discussed the results. M.G., R.V., and C.D. supervised all work.

\paragraph*{Competing interests}
M.G. is a cofounder and shareholder of QuEra Computing. All other authors declare no competing interests.

\nocite{*}

\bibliography{CSPT}

\end{document}